# SELF-SIMILAR STRUCTURE OF THE UNIVERSE AND FUNDAMENTAL SENSE OF EXISTENCE OF LARGE AND SMALL BLACK HOLES IN THE NATURE

**I.A. Kuchin, S.S. Boichenko, Y.I. Kuchin**

*Institute of Physics and Technology MES RK, Almaty, kuchin@satsun.sci.kz.*

The origin of self-similar (according to Y.Kulakov) structure of the Universe is discussed from a position of the theory of dynamic systems (DS). A probable nature of the isomorphism of DS configurations of different levels is revealed. Nucleon DS configuration like black hole (BH) might be acquired by the last as a result of Hawking radiation of initial BH and serve further as a genome of the Universe development .

## 1. Introduction

Two opposite answers on a question on a fundamental nature of the world are known [1]. According to the first (which was submitted by A. Einstein) the world is uniform in its basis and various forms of its complexity have the common nature. Therefore it is possible (and it is necessary) to geometrize interactions between objects, and the appropriate theory should have fundamental meaning. It is unique and may be considered as a truth in final instance. For this reason Einstein pushed forward his program of unification of electromagnetic and gravitational fields and interpreted gravitation as a real effect of the curved space.

The second position was submitted by F. Anderson as the concept of the "polyfundamental" nature of the world and at first sight looks like a purely phenomenological approach. According to this concept there are many fundamental theories in physics, and each of them is responsible for a certain level of organization of the physical world, but none of them can be considered more fundamental than others. According to hierarchical complexity of the physical world there is a hierarchy of theories, each one being effective only in the field of its competence. Instead of one true theory one has infinite 'tower' of effective theories, set up one on another, easily explaining the outward thin gs , but leaving the ends in an actual infinity uncontrollable by experiment. One end lies in the area of elements of a microcosm permanently diminishing and recessing from the human being, while the another one is in the far past of the expanding and developing Universe. In such "polyfundamental" approach solving the fundamental questions of the science is beyond the anthropic level and practical use.

The purpose of this paper is a presentation of a new viewpoint, free from extremes of the two positions mentioned above. This viewpoint based on the ideas and methods of the theory of dynamic systems (DS) and provides transition from the linear paradigm to the nonlinear one. According to it, the world seems to be multilevel, but these levels are not fundamental and are embraced by one general theory, the theory of DS. Problems which appear in such an approach are discussed below.

## 2. System isomorphism and self-similarity of the Universe

It follows from the theory of dynamic systems (DS), that an y system is not an object uniquely defined, but it is rather an infinite network of the objects connected by certain transitions, fittings. Namely, systems of different level of complexity may be presented by the equations of a different kind and there is a certain hierarchy of mathematical transitions between them: it is possible to pass from the equations of one type to the equations of another with help of some obvious simplifying (or complicating) assumptions [2]. Physically these assumptions correspond to a change of the object properties (or evolution of structure) depending on conditions of its existence.

Medial position in this hierarchy of mathematical structures is occupied by autonomous DS, gradient systems and potential functions. The equations of kinetics, describing nonequilibrium processes in dissipative media, i.e. diffusion, heat transmission and mass exchange adjoin to them. Adding of certain functions describing interaction of elements of a medium to the right part of such equations, allows one to consider cases of self-organizing of dynamic structures (and systems) in these mediums. Thus operating forces may have quite various nature (potentials of Newton, Coulomb, Yukawa). In other words, externally similar (but different in essence) objects can exist as the basis of action of different forces. For example, atomic system of electrons and star system of planets, large and small black holes. Global application of principles of isomorphism or micro – mega analogies results in a holographic paradigm of the Universe where the part is similar to the whole, and the whole to the part.

According to [3] the Universe has self-similar structure, with eleven levels which differ from each other by a factor $(e^{e^e})^{\pm n} = 3,8 \bullet 10^6$ :

$n = -5$ Plank region $l_{Pl}$ ~$10^{-33}$ cm;     $n = +5$ Metagalaxies $l_{meta}$ ~$10^{+28}$ cm;

$n = -4$ leptoquarks $l_{lq}$ ~$10^{-27}$ cm;     $n = +4$ galaxies $l_{gal}$ ~$10^{+24}$ cm;     (1)

$n = -3$ «gauge desert» $l_{gd}$ ~$10^{-21}$ cm;     $n = +3$ «space desert» $l_{sd}$ ~$10^{+18}$ cm;

$n = -2$ quarks $l_q$ ~$10^{-15}$ cm;     $n = +2$ planetary systems $l_{pl}$ ~$10^{+12}$ cm;

$n = -1$ atoms and molecules $l_{at}$ ~$10^{-9}$ cm;     $n = +1$ geological structures $l_{gs}$ ~$10^{+6}$ cm.

$n = 0$ objects of biosphere $10^{-5}$ cm $< l_{hum} < 10^{+2}$ cm.

This structure corresponds to the "allotropic" model of non-uniform growth of structures of different nature. Range of action of this parametrization extends from $10^{-33}$ up to $10^{+28}$ cm, with accuracy $\delta = 1\% - 5\%$. As it is known, allotropy expresses hierarchy of development processes and is widely represented in nature. It can be easily parametrized, but cannot be interpreted uniquely. Depending on a context one can speak about change of a system growth mode at different stages of development, or about change of parameters of medium (viscosity or inertness). In the first case one resorts to the help of multitime formalism, in the second the time of delay of system's reaction on external influence [4] enters. One can think that both interpretations are equivalent.

Complexity of big systems does not allow to study their nature in details, like solutions of Boltzmann or Yang-Mills equations. The behavior of complex objects is usually analyzed with the help of modelling, and even more complex can be analyzed step by step, by creation of some generalizing logic constructions on the basis of which the global conceptual theory is formed as a result. It is interesting, that key parameters of the world and main principles of its organization can be found with help of logic operations only (L.Korochkin). It allows biologists to consider living objects as certain fragments of general (cosmological) organization of the world.

The world of systems is not a conglomerate of the casual, disordered objects. On the contrary, the Universe as a whole acts as a certain self-similarly structured medium (see (1)) of objects of various nature, degree of complexity and size. Each level and each element of it obeys the universal allotropic dependence. The latter also concerns the level of elementary particles which appear to be not point-like, unstructured objects, but dynamic systems of the components which mass-spectrum satisfies the equation of the non-uniform growth of structures [5].

### 3. Nucleon as a big DS

It turned out, that nucleons and nucleus are arranged as «big systems». Despite of small size (diameter $10^{-15}$ cm), nucleon have all attributes of a big system. It has initial elements (quarks), there are forces of interaction between them (gluons), there is a law of a composition of system (configuration of the system) and there is the system "background", i.e. a certain motion of elements of lower level (partons). In this context the difference in sizes of the hadron constituents gets a fundamental value. In terms of Compton's wavelengths ($l_i = \hbar / m_i c$) their masses line up in a hierarchy of scales, characteristic for open systems in general according to Klimontovich. For typical hadron components ($t$-, $b$-, $c$-, $s$- etc. quarks) the following consequence takes place:

$$.. < l_H < l_t < l_b < l_c < l_s \leq L \approx 1\, fm, \qquad (2)$$

which is associated with the non-uniform growth of structures in the allotropic model [5].

The mass M growth in this model is described by the following equation

$$dM/dt = k(t)[t - \tau(t)]. \qquad (3)$$

Its solution is $M = M_0 \exp(zt)$, or $\ln(m) = zt$, here $m = M/M_0$, k – is the characteristic rate of growth, $\tau$ is the delay of the growth time, and z may be complex [4].

The macroscopic concept of medium viscosity can be concretized by certain quantum numbers, having connected them with different intensity of finite motion of components. As the motion of quarks is limited to the volume of a nucleon, its ground level can be approximated by certain circular (orbital) motion, and excitation can be presented by additional degrees of freedom. We mean oscillations on a way of this motion or rotation around the trajectory axis due to the spin. Intensity of oscillation can be characterized by the number of nodes $\nu$ in an orbit, intensity spiral motion by the "winding number" $\omega$. According to this, the formula (3) can be rewritten as

$$\ln m = f_1(\nu) + f_2(\omega), \qquad (4)$$

where $f_1(\nu)$ and $f_2(\omega)$ are contributions of the first and second kind of motion respectively. Concrete form of these contributions was studied in [6] on the basis of the model, submitted ad hoc by R. Milliken to explain the observable masses of leptons. His initial representation is a light front soliton, moving with the light speed $v = c$ in a certain closed space-time area. Authors [6] have found, that masses of e, $\mu$, $\tau$ leptons, bosons and quarks in correlation with $\nu$ and $\omega$ could be described by the following formula:

$$\ln m = -\frac{1}{2\pi\alpha}\left(\frac{1}{2}\right)^\omega + \ln\left(\frac{m_W}{\pi}\right) \quad \text{or} \quad \ln\left(\frac{m\pi^2}{\pi^\nu}\right) = -\frac{1}{2\pi\alpha}\left(\frac{1}{2}\right)^\omega + \ln(m_W) \qquad (5)$$

without any fitting parameters. Here $m_W$ is the mass of the W-boson, $\alpha = 1/137$, and $\nu$, $\omega$ are integers (1, 2, 3…).

### 4. Role of the DS configuration

Significance of a microsystem configuration reveals in properties of a macrosystem formed on its basis. Molecules of polymers illustrate "infinitely long" systems, fibers. Mesh polymers, crests are represented by rubbers. Endoedral structures are exemplified by fullerens, nanotubes and other objects.

Among this configuration variety the *DS* of a nucleon is an example of an "infinitely deep" system. Its diameter $D$ (=1,3 fm) is controlled by the mass of $\pi$-meson ($\cong 140$ *MeV*, $1/m_\pi \cong 1.4$ *fm*), and the depth by massive bosons, thousand times more massive ($m_H \geq 100$ *GeV*). A concrete image of such a configuration is a black hole. "Infinitely long" systems (fibers) and "infinitely deep" systems (hadrons) multiply by replicants and quarks never take off from a nucleon! In view of the characteristic size of the scale of Grand Unification ($l_{GU} \approx 3,8 \cdot 10^{-23}$ cm) and the Plank length of the formation ($l_{Pl} \approx 1,6 \cdot 10^{-33}$ cm) inequality (2) can be continued to the left:

$$\ldots < l_{Pl} < l_{GU} < l_H < l_t < l_b < l_c < l_s \leq L \approx 1\ fm, \text{ where } 1\ fm = 10^{-13} \text{ см}, \qquad (6)$$

and to be laced it with the hierarchy of lengths of open macrosystems by Klimontovich on the right. One gets the same hierarchical structure as in (2), but already inside of a nucleon.

A variety of properties of a nucleon DS on different depths of immersing (at different impact parameters) allows a nucleon to prove in various qualities, remaining itself and keeping the information included in it, and this has an important cosmological meaning [7]. Stratified, spherically-symmetric configuration of a nucleon DS promotes the reversion of the hadron wave front, getting in a nucleon, and it causes the Raman and induced Raman scattering (a breaks of the diffraction cone slope $B$ from *pp*-elastic scattering and ring events) [8]. In the case of $\pi p-$ and *pp* - interactions ring events look differently.

### 5. Micro-mega analogy

Large black holes (BH) are formed during a collapse of massive stars with masses exceeding the mass of the sun by tens and more times. At an early stage of development of the Universe "original" black holes with various masses (including microscopic) could be born. As a black body they might radiate and this radiation should consist of elementary particles accoring to Hawking. According to Hawking, radiation of BH emerges on the basis of crossing of principles of RT and the quantum mechanics represents fundamental interest for the theory. (Daghigh R.G.)

Now many physicists hope to observe BH at a level of nucleons and nucleus in experiments with particles of cosmic rays (Anchordoqui L.A.) or at a big hadron collider (Rizzo T.G.). In this area a real boom is observed in high energy physics. Actually a new physics beyond standard model of strong and electroweak interactions (Landsberg G.) is created here. Such factors as multidimensionality of space-time (Anchordoqui L.A.), existence of gravitational interaction between particles at small distances, feature of potential at $r \to 0$ (Shabad A.E.) and so on may result in formation of microholes.

The fundamental issue of the micromega analogy is to reveal a system-dynamic unity of the Universe.

### 6. Conclusions

From our point of view the essence of the affair is not a choice between mono - and polyfundamentalism, reductionism or antireductionism. They make sense only in a linear paradigm, where it is possible to disconnect the whole into parts. In the nonlinear approach the element is integrity of the object. At concentration and intensity of our influence on system, the letter acts either as compound object (the distributed system, or dissipative medium etc.), or as an indivisible whole (elementary essence, a substance, "atom" etc.) *Fundamentality is relative.*

At the level of linear approach (approximation) reductionism dominates and the main thing is the question on what all consists of and then this explains everything. We analyze a subsystem from a position supersystem. In the nonlinear approach the question on laws of the organization of complexities stands in the centre of attention. It one studies a supersystem on the basic of its subsystems. Functionality of the supersystem becomes informative and we search for what organizes complexity interesting for us.

As laws of dynamics are universal, it is not so essential to know, how the nature is arranged there where it is not accessible now to our experiment: at the moment of the big bang or in microcosm depths. It is important to ultimately understand its organization there where laws of systems self-organization are most accessible to the detailed analysis, i.e. at the anthropic level. Integrity of objects gives a key to understanding the nature of their components. Stable existence of a certain integrity is provided by certain properties of its components it, and the structure of the last by qualities of their elements etc. It is well-known, that values of physical constants provide a stable existence of the Universe in its present state, and the slightest deviation from them would be catastrophical.

## References

[1] Isaev P., Mamchur E. // Usp.Fiz. Nauk, **170** (2000) № 9, pp.1025-1030.

[2] Gilmore R. – Catastrophe theory for scientists and engineers, Virginia, 1981.

[3] Kulakov Yu.I.- About self-similar structure of the Universe // Proc. II Intern. conf."Selforganizing natural, technnogen. and social systems".Alma-Ata, sept.1998, p. 46-50.

[4] Zhirmunskij A.V., Kuz'min V.I. – Critical levels of development of natural systems. L., 1990.

[5] Kuchin I.A. – The dynamic structure of proton as open Klimontovich system. // Proc. Intern.Workshop "Problems of evolution of open systems", Almaty, 4-8 Oct., 1999, v. I, p.40.

[6] Millikan R.C., Richman D.C. – On the Masses of the Lepton, Bosons and Quarks. // arXiv: hep-th/0106106, 13 Jun 2001.

[7] Kuchin I.A., Boichenko S.S. – Proton as probable realization of the Universe genome concept. // Vestnik MNIIKA (Novosibirsk), № 10, 2004.

[8] Kuchin I.A. –To the question of hadron analogue of Raman light scattering. // Izvestia AN RK, (Almaty) № 2, 2003.